\documentclass{article}
\usepackage[dvips]{graphicx}
\usepackage{amssymb}
\usepackage{amsmath}
\usepackage{amscd}
\usepackage{mathrsfs}
\usepackage{latexsym}
\usepackage{makeidx}

\newtheorem{definition}{Definition}

\newtheorem{proposition}{Proposition}

\newcommand{\bd}{\begin{definition}}
\newcommand{\ed}{\end{definition}}
\newcommand{\bp}{\begin{proposition}}
\newcommand{\be}{\begin{equation}}
\newcommand{\ee}{\end{equation}}
\newcommand{\bea}{\begin{eqnarray}}
\newcommand{\eea}{\end{eqnarray}}
\newcommand{\ba}{\begin{array}}
\newcommand{\ea}{\end{array}}

\date{}
\author{Fabio Masillo,\footnote{E-mail address: masillo@le.infn.it} \, Giuseppe Scolarici,\footnote{E-mail address: scolarici@le.infn.it} \, Sandro Sozzo\footnote{E-mail address: sozzo@le.infn.it} \\ Dipartimento di Fisica and Sezione INFN \\ Universit\`a del Salento, via Arnesano, 73100 Lecce, Italy}

\title{\Large{\textbf{Proper Versus Improper Mixtures: Towards a Quaternionic Quantum Mechanics}}}
\begin{document}
\maketitle
\begin{abstract}
\noindent
The density operators obtained by taking partial traces do not represent \emph{proper mixtures} of the subsystems of a compound physical system, but \emph{improper mixtures}, since the coefficients in the convex sums expressing them never bear the ignorance interpretation. As a consequence, assigning states to these subsystems is problematical in standard quantum mechanics (\emph{subentity problem}). Basing on the proposal provided in the SR interpretation of quantum mechanics, where improper mixtures are considered as true nonpure states conceptually distinct from proper mixtures, we show here that proper and improper mixtures can be represented by different density operators in the quaternionic formulation of quantum mechanics, hence they can be distinguished also from a mathematical viewpoint. A simple example related to the quantum theory of measurement is provided.

%\vspace{.3cm}
%\noindent 
%\textbf{Key Words:} quaternionic quantum mechanics; proper and improper mixtures; subentity problem; SR interpretation. 
\end{abstract}

\section{Introduction}
In the complex formulation of quantum mechanics (CQM) a physical system $\Omega$ is associated with a separable complex Hilbert space ${\mathscr H}^{{\mathbb C}}$ and the states of $\Omega$ are represented by density operators on ${\mathscr H}^{{\mathbb C}}$, which reduce to one-dimensional (orthogonal) projection operators in the case of \emph{pure} states. Every density operator $\rho$ representing a \emph{mixed} state, or \emph{proper mixture}, $S_M$ of $\Omega$ can be expressed in many ways as a convex combination of pure states, and a decomposition $\rho=\sum_{i} p_{i}| \psi_{i} \rangle\langle \psi_{i}|$ exists in which every coefficient $p_i$ denotes the probability that $\Omega$ be in the state $S_i$ represented by the projection operator $| \psi_{i} \rangle\langle \psi_{i}|$. A proper mixture can be produced by performing a (nonselective) measurement, thus obtaining the wave function collapse which is a nonlinear process in quantum mechanics. The probability $p_i$ expresses our ignorance about the real state of $\Omega$, hence also about the result of a measurement testing whether the property $E_i$ of $\Omega$ represented by $|\psi_{i}\rangle\langle\psi_{i}|$ is possessed by $\Omega$. Yet, if $\langle\psi_i|\psi_j \rangle=\delta_{ij}$, every $E_i$ is \emph{objective} in $S_M$, in the sense that it can be considered as either possessed or not possessed by $\Omega$ independently of any measurement. 

Let now $\Omega$ be a compound system, made up of two subsystems $\Omega_1$ and $\Omega_2$, prepared in a pure entangled state $S_P$ represented by the projection operator $| \psi \rangle\langle \psi|$. Let $|\psi\rangle=\sum_{i}\sqrt{p_i}|\phi_{i}(1)\rangle |\chi_{i}(2)\rangle$$, 0 < p_i < 1$, be the biorthogonal decomposition of $|\psi\rangle$. If one considers $\Omega_1$ only, the physical information provided by CQM on it can be attained by taking the partial trace of $|\psi\rangle\langle\psi|$ with respect to $\Omega_2$, thus getting $\rho_1= Tr_{2} |\psi \rangle\langle \psi|=\sum_{i} p_{i} |\phi_{i}(1)\rangle\langle\phi_{i}(1)|$. The density operator $\rho_1$ is formally similar to $\rho$. Yet, a coefficient $p_i$ in it denotes the probability of actualizing the property $E_i(1)$ of $\Omega$ represented by the projection operator $|\phi_{i}(1)\rangle\langle\phi_{i}(1)| \otimes I_2$ whenever a measurement occurs, but it cannot denote the probability that $\Omega_1$ actually be in the state $S_{i}(1)$ represented by $|\phi_{i}(1)\rangle\langle\phi_{i}(1)|$. Indeed, $E_i(1)$ should then be objective, as the property $E_i$ considered above, while it is \emph{nonobjective} in $S_P$ according to the standard interpretation of quantum mechanics (that is, one cannot consider $E(1)$ as either possessed or not possessed by $\Omega$ in the state $S_P$ if a measurement is not performed).\footnote{Nonobjectivity is commonly believed to be an intrinsic and uneliminable feature of standard quantum mechanics because of some mathematical results, as the \emph{Bell-Kochen-Specker theorem} \cite{b66,ks67} and the \emph{Bell theorem} \cite{b64}. Yet, it is the deep root of most problems that afflict the standard interpretation and raises a lot of paradoxes and conceptual difficulties (in particular, the \emph{objectification problem} in the quantum theory of measurement, see, \emph{e.g.}, \cite{blm91,bs96}). \label{nogo}} Basing on this conclusion, one can show that no decomposition of $\rho_1$ bears the above \emph{ignorance interpretation}. Hence, some authors say that $\rho_1$ represents an \emph{improper mixture}, distinguishing it from a proper mixture as $\rho$ (see, \emph{e.g.}, \cite{hk69}--\cite{blm91}). As a consequence, the density operators obtained by taking partial traces generally neither represent pure nor mixed states of the component subsystems in standard quantum mechanics, so that these subsystems can never be considered as independent entities, which raises the so-called \emph{subentity problem} \cite{a99a}--\cite{a99b,a00}. 

A proposal of solution of the subentity problem has recently been forwarded by one of the authors \cite{gs06} in the framework of the \emph{Semantic Realism} (or \emph{SR}) interpretation of quantum mechanics \cite{gs96a}--\cite{ga02}. The SR interpretation has been worked out to show that the mathematical apparatus of quantum mechanics is compatible with objectivity of physical properties, which avoids the objectification problem together with a number of quantum paradoxes. In this perspective, improper mixtures are considered as new states of the physical system and they cannot be distinguished in CQM from proper mixtures. Yet, the unitary evolution of proper mixtures is different from the evolution of improper mixtures, generally nonunitary, which suggests, according to this interpretation, a more suitable mathematical representation of these physically different entities.

In 1936, by using lattice theoretic arguments, Birkhoff and von Neumann \cite{bn36} concluded that the set of states of a quantum system can be represented by a vector space over the real $\mathbb{R}$, complex $\mathbb{C}$, or quaternionic $\mathbb{Q}$, fields. While the real number formulation of quantum mechanics is essentially equivalent to CQM \cite{st60}, the research on quaternionic quantum mechanics (QQM) began much later with a series of papers by Finkelstein \emph{et al.} in the sixties \cite{fjss60}, and pursued up to now. A systematic study of QQM is given in  \cite{a95}.

The possibility of a generalization of quantum mechanics based on quaternion fields instead of complex fields is still controversial. Yet, the rich structures emerging from such a generalization proved to be useful in the description of entanglement, dynamical maps and decoherence phenomena in quantum physics \cite{Kossa1}--\cite{as}. Furthermore, we will see in this paper that these structures provide rigorous tools to solve the subentity problem.

Let us briefly sketch the content of the next sections.

After briefly resuming the qualitative solution of the subentity problem that raises from the SR interpretation of quantum mechanics in Sec. 2, we will discuss in Sec. 3 how mixtures are represented in QQM. In particular, we will show that proper and improper mixtures can be represented by different density operators in QQM, and that this mathematical representation is compatible with their different time evolutions in CQM. Finally, we will apply these results in Sec. 4, and consider a specific example regarding the measurement process.

\section{The subentity problem in the SR interpretation of quantum mechanics \label{sr}}
The SR interpretation of quantum mechanics \cite{gs96a}--\cite{ga02} has been worked out in order to show that both CQM and QQM can be embodied into a more general framework in which objectivity of physical properties holds and quantum probabilities are reinterpreted as conditional instead of absolute. The SR interpretation avoids the objectification problem and other quantum paradoxes. Moreover, it provides a nonstandard solution of the subentity problem \cite{gs06}, which will be summarized in the following.

(i) A physical system $\Omega$ is associated with a set $\mathcal S$ of \emph{states} and a set $\mathcal E$ of \emph{physical properties}. Each state is operationally defined as a class of statistically equivalent \emph{preparing devices}. Each property is operationally defined as a class of statistically equivalent \emph{ideal registering devices}.

(ii) We assume that all properties in $\mathcal E$ are \emph{objective}, \emph{i.e.}, for every $E \in \mathcal E$, the outcome of a registering device $r \in E$, when applied to an individual sample (or, \emph{physical object}) $x$ of $\Omega$, does not depend on the measurement procedure.\footnote{Objectivity of properties implies that the SR interpretation clashes with the standard interpretation, which asserts instead nonobjectivity of properties on the basis of empirical (\emph{e.g.}, the double-slit experiment) or theoretical (\emph{e.g.}, the no-go theorems mentioned in footnote \ref{nogo}) arguments. Hence, the SR interpretation was worked out together with an accurate analysis of those arguments, which singled out some weaknesses in each of them. In particular, theoretical arguments in favor of nonobjectivity turn out to be based on implicit assumptions that, when made explicit, are rather doubtful. Indeed, these assumptions subtend an epistemological perspective that assumes the validity of empirical quantum laws also in physical situations in which quantum mechanics itself states that, in principle, they cannot be checked \cite{gs96b}--\cite{ga00}. If this perspective is criticized, nonobjectivity appears as an interpretative choice, not a logical consequence of the theory, and alternative objective interpretations (as the SR) become possible.}

(iii) It follows from (ii) that the probability of finding a given result when performing a measurement on a physical object $x$ can be interpreted as expressing our ignorance about the properties possessed by $x$ (in this sense one can say that it is \emph{epistemic}) in the SR interpretation, whatever the state of the physical object may be. The distinction between pure and nonpure states may still be introduced basing on the different values of the probabilities of the properties in these states, but not on different interpretations (epistemic or not) of the probabilities themselves. In particular, one can accept the standard representation of states by means of density operators, and characterize pure states as the states whose representing density operators reduce to projection operators.

(iv) Since every state is operationally defined as an equivalence class of preparing devices, if one considers a state $S$ represented by the density operator $\sum_{i} p_i |\psi_i\rangle\langle\psi_i|$, an ensemble of physical objects in the state $S$ can be realized by a \emph{mixed} preparing device, \emph{i.e.}, a device that mixes physical objects prepared by devices belonging to the states $S_1$, $S_2$, \ldots represented by the projection operators $|\psi_1\rangle\langle\psi_1|$, $|\psi_2\rangle\langle\psi_2|$, \ldots, respectively. In this case a coefficient $p_i$ cannot only be interpreted as in (iii), but also as the probability that a given physical object in the state $S$ actually be in the state $S_i$. Nevertheless, there is no evident physical reason, according to the SR interpretation, for assuming that $S$ contains only mixed preparing devices. 

(v) It follows from (iv) that in the case of compound physical systems the density operators obtained by taking partial traces can be accepted as representing states in which also preparations occur that are not mixed in the sense specified in (iv).

(vi) The probabilistic definition of states in (i) groups together, in the case of nonpure states, mixed with nonmixed preparing devices, that therefore cannot be distinguished by means of measurements. This opens the way to a possible solution of the problem of explaining how both unitary and nonunitary evolutions may occur for the same density operator, since it suggests distinguishing mixed from nonmixed preparing devices by introducing a new equivalence relation on the set of all preparing devices, strictly contained in the physical equivalence relation defined in (i). Thus, every state $S$ would be associated with a family of \emph{hidden states}, which would be equivalent with respect to measurements but could have different behaviours with respect to time evolution.

\section{Mixtures in quaternionic quantum mechanics}
We recall in the first part of this section some basic notations, properties and results of QQM (for an exhaustive discussion of quaternionic matrices, see, \emph{e.g.}, \cite{Zhang}) in order to provide in the second part different mathematical representations for proper and improper mixtures. 

%A (real) quaternion is usually
%expressed as 
%\begin{displaymath}
%q=q_{0}+q_{1}i+q_{2}j+q_{3}k
%\end{displaymath}
%where $q_{l}\in \mathbb{R}$ ($l=0,1,2,3$), $i^{2}=j^{2}=k^{2}=-1,$ $ij=-ji=k$.

%The quaternion skew-field $\mathbb{Q}$ is an algebra of rank $4$ over $\mathbb{R}$, noncommutative and endowed with an involutive anti-automorphism (\emph{conjugation}) such that 
%\begin{displaymath}
%q \rightarrow \bar{q}=q_{0}-q_{1}i-q_{2}j-q_{3}k.
%\end{displaymath}
%In a (right) $n$-dimensional vector space $\mathbb{Q}^{n}$ over $\mathbb{Q}$, every linear operator is associated in a standard way with a $n\times n$ matrix acting on the left. Moreover, in analogy with the case of vector spaces over $\mathbb{C}$, one can introduce the concepts of unitarity, hermiticity and so on (for the sake of simplicity, we will limit ourselves to consider only finite-dimensional quaternionic Hilbert spaces; the generalization to infinite dimension is straightforward).

A physical system $\Omega$ is associated in QQM with a quaternionic $n$-dimensional right Hilbert space ${\mathscr H}^{{\mathbb Q}}$ \cite{a95} (for the sake of simplicity, we will limit ourselves to consider finite-dimensional quaternionic Hilbert spaces; this will allow us to denote operators and the associated matrices by the same symbols). The states of $\Omega$ are represented by positive hermitian operators on ${\mathscr H}^{{\mathbb Q}}$ with unit trace (as in CQM). More precisely, a pure state $S_P$ of $\Omega$ is represented by a density operator $\rho_{\psi}=|\psi \rangle \langle \psi |$ (where $|\psi\rangle$ is a unit vector of ${\mathscr H}^{{\mathbb Q}}$) with rank one, while a mixed state $S_M$ of $\Omega$ is represented by a density operator $\rho$ with rank greater than one.

The observables of $\Omega$ are represented by hermitian operators on ${\mathscr H}^{{\mathbb Q}}$. Moreover, the expectation value of an observable $\mathcal A$, represented by the quaternionic hermitian operator $A$, in the pure state $S$, represented by the unit vector $|\psi\rangle$, is given by \cite{a95}
\begin{equation} \label{expec}
\langle A \rangle_{\psi} = \langle \psi |A|\psi \rangle =\textrm{Re} \textrm{Tr} (A|\psi \rangle \langle \psi |)=\textrm{Re} \textrm{Tr} (A\rho_{\psi}). 
\end{equation}
Expanding $A=A_{\alpha}+jA_{\beta}$ and $\rho=\rho_{\alpha}+j\rho_{\beta}$ (where $j$ denotes one of the quaternion imaginary units) in terms of the complex matrices $A_{\alpha}$, $A_{\beta}$, $\rho_{\alpha}$ and $\rho_{\beta}$, it follows that the expectation value $\langle A\rangle_{\psi}$ may depend on $A_{\beta}$ or $\rho_{\beta}$ only if both $A_{\beta}$ and $\rho_{\beta}$ are different from zero. Indeed,
\begin{equation} \label{esplexpc}
\langle A\rangle_{\rho}=\textrm{Re}\textrm{Tr}(A\rho)=\textrm{Re}\textrm{Tr}(A_{\alpha}\rho_{\alpha}-A_{\beta}^{\ast}\rho_{\beta}),
\end{equation}
where $\ast $ denotes complex conjugation. Thus, the expectation value of an observable $\mathcal A$, represented by the hermitian operator $A$, in the state $S$, represented by the density matrix $\rho$, depends on the quaternionic parts of $A$ and $\rho $ only if both the observable and the state are represented by genuine quaternionic matrices. However, if an observable $\mathcal A$ is represented by a pure complex hermitian matrix, its expectation value in the state $S$ does not depend on the quaternionic part $j\rho _{\beta }$\ of the density matrix $\rho =\rho _{\alpha}+j\rho_{\beta}$ representing $S$. Moreover, the expectation value predicted in CQM in the state represented by the density matrix $\rho_{\alpha}$ coincides with the expectation value in the state represented by $\rho$ predicted by QQM, since 
\begin{equation}
\textrm{Tr}( A \rho_{\alpha})=\textrm{Re} \textrm{Tr}(A \rho_{\alpha})=\textrm{Re}\textrm{Tr}(A\rho).
\end{equation}
This simple observation is relevant in our approach since it enables us to merge CQM in the (more general) framework of QQM without modifying any theoretical prediction (as long as complex observables are taken into account), thus eluding or postponing any comparison between these formulations.

Let us now denote by $M(\mathbb{Q})$ and $M(\mathbb{C})$ the space of $n \times m$ quaternionic and complex matrices, respectively, and let $M=M_{\alpha}+j M_{\beta} \in M(\mathbb{Q})$. We define the \emph{complex projection}
\begin{displaymath}
P: M(\mathbb{Q}) \rightarrow M(\mathbb{C})
\end{displaymath}
by the relation
\begin{equation} \label{complexProjection}
P(M)=\frac{1}{2}[M-i M i]=M_{\alpha}.  
\end{equation}

When we consider time-dependent quaternionic unitary dynamics,
\begin{equation}
\rho (t)=U(t)\rho (0)U^{\dagger }(t),  \label{unitev}
\end{equation}
where
\begin{equation}
U(t)=(U_{\alpha }+jU_{\beta })(t)=T_{o}\,e^{-\int_{0}^{t}duH(u)}
\label{timeordev}
\end{equation}
and $T_{o}$ denotes the time ordering operator, the differential equation associated with the time evolution for $\rho $ reads
\begin{equation}
\frac{d}{dt}\rho (t)=-[H(t),\rho (t)],  \label{eveqtimdep}
\end{equation}
where $H(t)=H_{\alpha }+jH_{\beta }=-\left( \frac{d}{dt}U(t)\right) U^{\dagger }(t)$. Finally, Eqs. (\ref{unitev}) and (\ref{eveqtimdep}) reduce to
\begin{equation}
\rho _{\alpha }(t)=U_{\alpha }\rho _{\alpha }(0)U_{\alpha }^{\dagger
}+U_{\beta }^{\ast }\rho _{\alpha }^{\ast }(0)U_{\beta }^{T}+U_{\alpha }\rho
_{\beta }^{\ast }(0)U_{\beta }^{T}-U_{\beta }^{\ast }\rho _{\beta
}(0)U_{\alpha }^{\dagger }  \label{finitecomppro}
\end{equation}
and
\begin{equation}
\frac{d}{dt}\rho _{\alpha }=-[H_{\alpha },\rho _{\alpha }]+H_{\beta }^{\ast
}\rho _{\beta }-\rho _{\beta }^{\ast }H_{\beta },  \label{gencompro}
\end{equation}
respectively, for the complex projection of the density matrix \cite{Asor}.

Now, we focus our attention on the complex projection $\rho _{\alpha }$ of a quaternionic density matrix $\rho =\rho _{\alpha }+j\rho _{\beta }$.

First of all, it follows from the hermiticity of $\rho$ and $\rho_{\alpha}$ that
\begin{displaymath}
\textrm{Tr}\rho_{\alpha}=\textrm{Re}\textrm{Tr} \rho_{\alpha}=\textrm{Re}\textrm{Tr}\rho=\textrm{Tr}\rho,
\end{displaymath}
\emph{i.e.}, the complex projection of any quaternionic density matrix is trace preserving. Moreover, we recall that \cite{Asor}:

\vspace{.2cm}
\noindent
\textbf{Proposition 1.} \emph{The complex projection $\rho_{\alpha }$ of any quaternionic density matrix $\rho=\rho_{\alpha}+j\rho_{\beta}$ is a complex density matrix}. 

\vspace{.2cm}
\noindent
The following statement provides instead informations about the rank of the complex projection $\rho _{\alpha }$ of any quaternionic density matrix $\rho =\rho_{\alpha}+j\rho_{\beta}$ \cite{as}.

\vspace{.2cm}
\noindent
\textbf{Proposition 2.} \emph{Let $\rho=\rho_{\alpha}+j\rho_{\beta}$ be a $n$-dimensional quaternionic density matrix, and let $\mathrm{rank}\,\rho =m$. Then, $m$ $\leq \mathrm{rank}\,\rho_{\alpha}\leq 2m$.}

\vspace{.2cm}
\noindent
Conversely \cite{gallipoli},

\vspace{.2cm}
\noindent
\textbf{Proposition 3.} \emph{Let $\rho_{\alpha}$ be a $n$-dimensional complex density matrix with $\mathrm{rank}\,\rho
_{\alpha }=m>1$, and let $[x]$ denote the integer part of $x$. Then, for any $m^{\prime}$ with $\left[ \frac{m+1}{2}%
\right] $ $\leq m^{\prime }\leq m$ there exists a (skew-symmetric) complex matrix $\rho _{\beta }$ such that $\rho =\rho _{\alpha}+j\rho _{\beta }$ is a density matrix with $\mathrm{rank}\,\rho=m^{\prime }$.}

\vspace{.2cm}
\noindent
As a consequence of the above two propositions, we can conclude that:

\vspace{.2cm}
\noindent
\textbf{Proposition 4.} \emph{Any complex density matrix $\rho _{\alpha }$ can be obtained as the complex projection of a quaternionic pure density matrix $\rho =\rho _{\alpha }+j\rho _{\beta }$ if and only if $\mathrm{rank} \rho _{\alpha }=2$}.

\vspace{.2cm}

Let us now discuss how the above results can be used to represent proper and improper mixtures by different density matrices in QQM. To this aim, we observe that every complex density matrix $\rho_\alpha$ can be associated with a set $[\rho_\alpha]$ of quaternionic density matrices as follows: 
\begin{displaymath}
\rho_\alpha\longrightarrow[\rho_\alpha]=\{\rho=\rho_\alpha+j\rho_\beta\},
\end{displaymath}
where the $\rho_{\beta}$s must be chosen in a such way that $\rho$ is still a density matrix (see Propositions 1--4).

The above mapping can be inverted by means of the complex projection in Eq. (\ref{complexProjection}), and it introduces an equivalence relation $\approx$ on the set of density matrices defined, for every $\rho, \rho\prime$, as $\rho \approx \rho\prime$ iff $P(\rho)=P(\rho\prime)$. Hence, the set of quaternionic density matrices is partitioned into equivalence classes, and each class contains one and only one complex density matrix, which thus represents the class. 

Each equivalence class $[\rho_{\alpha}]$ can further be partitioned by distinguishing the quaternionic density matrices $\rho=\rho_{\alpha}+j\rho_{\beta}$, $\rho_{\beta} \ne 0$, from the complex matrix $\rho=\rho_{\alpha}$. It is important to observe that a complex unitary dynamics, \emph{i.e.}, a dynamics given by a complex unitary matrix $U_{\alpha}$, preserves both the first and the second partition. Indeed, we get from Eqs. (\ref{unitev}) and (\ref{finitecomppro}) that
\begin{equation}
\rho(t)=\rho_\alpha(t)+j \rho_\beta(t)=U_{\alpha}\rho_\alpha(0)U_{\alpha}^{\dagger}+j U_{\alpha}^{*}\rho_\beta(0)U_{\alpha}^{\dagger},
\end{equation}
\begin{equation}
\rho_\alpha(t)=U_{\alpha}\rho_\alpha(0)U_{\alpha}^{\dagger}.
\end{equation}
On the contrary, a quaternionic unitary dynamics in general neither preserves the first nor the second partition (see again Eqs. (\ref{unitev}) and (\ref{finitecomppro})).

As stated in the previous sections, the distinction between proper and improper mixtures is strictly connected in CQM with the existence of entangled states and the partial trace procedure. In particular, we notice that the latter operation can give rise to the following situations for the subsystems.

(i) A separable state of the compound system generally produces proper mixtures of the component subsystems.

(ii) An entangled state of the compound system generally produces improper mixtures of the component subsystems.
%\begin{eqnarray*}
%  \mbox{Separable State} &\longrightarrow&   \mbox{Proper Mixtures}, \\
%    \mbox{Entangled State} &\longrightarrow&   \mbox{Improper Mixtures}. 
%\end{eqnarray*}

Clearly, if the unitary evolution of the compound system is factorizable in CQM, it transforms entangled states into entangled states, and  separable states into separable states. Then we demand that, if a mathematical distinction between proper and improper mixtures exists in QQM, then the unitary subdynamics associated in CQM with each subsystem should be such that proper mixtures are transformed into proper mixtures and improper mixtures are transformed into improper mixtures. Therefore, we are led to represent proper mixtures by the quaternionic density matrices for which $P(\rho)=\rho$, and improper mixtures by the quaternionic density matrices  for which $P(\rho)\neq\rho$. Of course, a complex unitary dynamics does not modify this distinction. In fact, if $\rho_\beta(0)=0$, then $\rho_\beta(t)=U_{\alpha}^{*}\rho_\beta(0)U_{\alpha}^{\dagger}=0$.

We stress once again that the density matrices $\rho$ representing proper and improper mixtures produce the same expectation values on complex observables.

We finally observe that the above partition is not preserved in CQM by nonfactorizable dynamics of the compound system. In fact, in this case the subsystems can exchange entanglement each other during their evolution, which is thus nonunitary. We remind that nonunitary dynamics in CQM can in many cases be described in terms of the complex projection of a quaternionic unitary evolution \cite{compent,Asor}, \cite{as}.

\section{The measurement process: an illustrative example}
Our main aim in this section is to illustrate an example in which QQM represents proper and improper mixtures with different density operators, thus allowing one to distinguish them not only at a conceptual but also at a mathematical level. The proposed example is meaningful since it concerns with the description of the measurement process as a dynamical process and is highly problematical in standard quantum mechanics. 

We will consider a very simple and schematic model to describe the interaction between the measured system and the measuring apparatus that occurs in a measurement. More precisely, let $\Omega_m$ be a microscopic physical system, for instance, a spin-$\frac{1}{2}$ quantum particle, associated with the complex Hilbert space ${\mathscr H}_{m}^{\mathbb C}={\mathbb C}^{2}$, and let $\mathcal A$ be the observable ``spin of $\Omega_m$ along the direction $\vec{n}$'', represented in CQM by the hermitian operator $A=\frac{1}{2}\hbar \vec{\sigma}\cdot \vec{n}$, where $\vec{\sigma}$ are the Pauli matrices. Let $s_1$ and $s_2$ be the pure states of $\Omega_m$ corresponding to the eigenvectors $| +_{\vec{n}} \rangle$ and $| -_{\vec{n}} \rangle$ of $A$, respectively. Let us schematize the apparatus that performs an ideal measurement of $\mathcal A$ on $\Omega_m$ by means of a macroscopic physical system $\Omega_M$ that can be described in CQM by the Hilbert space ${\mathscr H}_{M}^{\mathbb C}={\mathbb C}^{2}$. Suppose that $\Omega_M$ is initially in the macroscopic state $S_0$ represented by the unit vector $|0 \rangle$ corresponding to the value 0 on the reading scale of the apparatus, and that $\Omega_M$ possesses further macroscopic states $S_1$ and $S_2$ represented by the unit vectors $|u_{\vec{n}} \rangle$ and $| d_{\vec{n}} \rangle$, respectively (corresponding to the values $up$ and $down$, respectively, on the same scale), and let $\{|u_{\vec{n}} \rangle,|d_{\vec{n}} \rangle  \}$ be an orthonormal basis on ${\mathscr H}_{M}^{\mathbb C}$. Furthermore, assume that there exists a one-to-one correspondence between the states $s_1$ and $s_2$ of $\Omega_{m}$ and the states $S_1$ and $S_2$ of $\Omega_{M}$. Let $s_0$ be the initial state of $\Omega_m$ represented by the unit vector $|\varphi_0\rangle=c_+ |+_{\vec{n}} \rangle + c_- |-_{\vec{n}} \rangle$. Finally, assume that the interaction between $\Omega_m$ and $\Omega_M$ is represented by a complex unitary operator $U(0,t)$. This interaction can be described as follows.
\begin{equation} \label{unitcomp}
\begin{CD}
| \Psi(0)\rangle=|\varphi_0\rangle |0\rangle =[c_+ |+_{\vec{n}} \rangle + c_- |-_{\vec{n}} \rangle] |0\rangle @>>U(0,t)> |\Psi (t)\rangle= c_+ |+_{\vec{n}} \rangle |u_{\vec{n}} \rangle+c_- |-_{\vec{n}} \rangle |d_{\vec{n}} \rangle, 
\end{CD}
\end{equation}
where $|\Psi(t)\rangle$ is expressed by a biorthogonal decomposition, hence it represents a pure entangled state of the compound system $\Omega_m+\Omega_M$. The physical information on $\Omega_m$ can be obtained in CQM by taking the partial trace of the density operator $\rho_{\alpha}(t)=|\Psi(t)\rangle\langle\Psi(t)|$ with respect to the physical system $\Omega_M$. Then, we get
\begin{equation} \label{partialtrace}
\rho_{\alpha}^{m}(t)=Tr_{M} \rho_{\alpha}(t)=\langle u_{\vec{n}} | \rho_{\alpha}(t) | u_{\vec{n}} \rangle+\langle d_{\vec{n}} | \rho_{\alpha}(t) | d_{\vec{n}} \rangle=|c_+|^{2} | +_{\vec{n}}\rangle\langle +_{\vec{n}}|+|c_-|^{2} | -_{\vec{n}}\rangle\langle -_{\vec{n}}| .
\end{equation}
This density operator represents an improper mixture $M_{S}^{I}$ of $\Omega_m$ (see Sec. 1). 

If we now consider the physical system $\Omega_m$ separately, and apply the projection postulate of CQM in the case of a nonselective measurement of the observable $\mathcal A$ we instead obtain, via L\"{u}ders' rule,
\begin{equation} \label{projpost}
\rho_{\alpha}^{m}=|+_{\vec{n}}\rangle\langle +_{\vec{n}}|\varphi_0\rangle\langle\varphi_0|+_{\vec{n}}\rangle\langle +_{\vec{n}}|+|-_{\vec{n}}\rangle\langle -_{\vec{n}}|\varphi_0\rangle\langle\varphi_0|-_{\vec{n}}\rangle\langle -_{\vec{n}}|=\rho_{\alpha}^{m}(t).
\end{equation}
This density operator formally coincides with the one obtained in Eq. (\ref{partialtrace}) but it represents a proper mixture $M_{S}^{P}$ of $\Omega_m$ in this case (see again Sec. 1). 

It is well known that the main problem of the standard quantum theory of measurement is reconciling the two above descriptions, in particular, recovering the objectification, that is, the occurrence of definite outcomes for the macroscopic apparatus (\emph{objectification problem}). Any attempt at providing a consistent description of the measuring process within the quantum formalism and its standard interpretation leads to the so-called \emph{von Neumann's chain} and to the ensuing quantum paradoxes (in particular, \emph{Schr\"{o}dinger's cat} and \emph{Wigner's friend} paradoxes).

We have observed in Sec. 2 that the above problem disappears in the SR interpretation of quantum mechanics because of objectivity of the properties of both $\Omega_m$ and $\Omega_M$. However, $S_{M}^{P}$ and $S_{M}^{I}$ cannot be distinguished in CQM also if the SR interpretation is adopted, since they are represented by the same complex density operator. Yet, we can use the arguments in Sec. 3 to accomplish this task in QQM.

According to the partition introduced in Sec. 3, the improper mixture $S_{M}^{I}$ can be represented in QQM by the density operator $\rho^{m}=\rho_{\alpha}^{m}+j \rho_{\beta}^{m}$, where $\rho_{\beta}^{m} \ne 0$, while the proper mixture $S_{M}^{P}$ can be represented by the density operator ${\rho}'^{m}=\rho_{\alpha}^{m}$. It must be noted that $S_{M}^{I}$ can also be purified\footnote{The possibility of considering improper mixtures as new pure states has already been taken into account by Aerts in a series of papers \cite{a99a}--\cite{a00} where, however, no mention to the field of the Hilbert space has been made.} (see Proposition 4) by choosing
\begin{equation}
\rho^{m}=|c_{+}|^{2}|+_{\vec{n}}\rangle\langle +_{\vec{n}}|+|c_{-}|^{2}|-_{\vec{n}}\rangle\langle -_{\vec{n}}|+|-_{\vec{n}}\rangle j c_{+}^{*}c_{-}^{*} \langle +_{\vec{n}}|-|+_{\vec{n}}\rangle j c_{+}^{*}c_{-}^{*} \langle -_{\vec{n}}|,
\end{equation}
which is the projection operator on the 1--dimensional subspace generated by the unit vector $|+_{\vec{n}}\rangle c_{+} +|-_{\vec{n}}\rangle c_{-}j$. One realizes at once that $\rho^{m}$ and ${\rho}'^{m}$ produce the same expectation values on complex observables, whereas their expectation values are different if purely quaternionic observables are taken into account.

The above result is relevant from our viewpoint since it shows that the mixtures $S_{M}^{I}$ and $S_{M}^{P}$ can be experimentally distinguished, at least in principle, in QQM.

\end{document}